\documentclass[runningheads]{svmult}
\usepackage{makeidx}   
\usepackage{graphicx}  
\usepackage{subeqnar}  
\usepackage{multicol}  
\usepackage{cropmark} 
\usepackage{physprbb}  
\def\ga{\mathrel{\raise.3ex\hbox{$>$\kern-.75em\lower1ex\hbox{$\sim$}}}}
\def\la{\mathrel{\raise.3ex\hbox{$<$\kern-.75em\lower1ex\hbox{$\sim$}}}}
\def\he#1{\hbox{${}^{#1}$He}}
\def\li#1{\hbox{${}^{#1}$Li}}
\def\be#1{\hbox{${}^{#1}$Be}}
\def\b#1#2{\hbox{${}^{#1#2}$B}}
\def\beq{\begin{equation}}
\def\eeq{\end{equation}}

\begin{document}
\title*{
\begin{flushright}
{\rm \small UMN-TH-1923/00 \\
TPI-MINN-00/47 \\
 CERN-TH/2000-287\\
astro-ph/0009475 \\
September 2000}
\end{flushright}
\vskip -1.2in
Big Bang Nucleosynthesis\protect\newline and  Related Observations}
\toctitle{Big Bang Nucleosynthesis and  Related Observations}
%
%
\titlerunning{Big Bang Nucleosynthesis}
%
\author{Keith A. Olive\inst{1,2,*}}
\authorrunning{Keith A. Olive}
%
%
\institute{TH Division, CERN, Geneva, Switzerland
\and Theoretical Physics Institute,
School of Physics and Astronomy,\\ University of Minnesota, Minneapolis MN,
USA \\
\hskip -.14in $^*$ ~To be published in the Proceedings of 
Dark 2000: Third International Conference on Dark Matter in
         Astro and Particle Physics,
Heidelberg, Germany, July 10-16 2000}

\maketitle              

\begin{abstract}
The current status of big bang nucleosynthesis is summarized.  Particular attention
is paid to recent observations of \he4 and  \li7 and their systematic uncertainties. Be
and B are also discussed in connection to recent \li7 observations and the primordial
\li7 abundance.
\end{abstract}

\section{Introduction}

The simplicity of the standard model of Big Bang Nucleosynthesis (BBN) and its 
success when confronted with observations place the theory as one of the cornerstones
of modern cosmology.
BBN
is based on the inclusion of an extended nuclear
network into a homogeneous and isotropic cosmology.  Apart from the
input nuclear cross sections, the theory contains only a single parameter,
namely the baryon-to-photon ratio,
$\eta$. Other factors, such as the uncertainties in reaction rates, and
the neutron mean-life can be treated by standard statistical and Monte
Carlo techniques\cite{kr,hata1,sark2,bn}.  The theory then allows one to make
predictions (with specified uncertainties) of the abundances of the light elements,
D, \he3, \he4, and \li7. As there exist several detailed reviews on BBN, I will
briefly summarize the key results and devote this contribution to the impact of
recent observations of \he4 and \li7 along with the related observations of Be
and B. In referring to the standard model, I will mean homogeneous nucleosynthesis,
with three neutrino flavors ($N_\nu = 3$), and a neutron mean life of 886.7 $\pm$
1.9 s \cite{RPP}.

\begin{figure}[ht]
\begin{center}
\includegraphics[width=25pc]{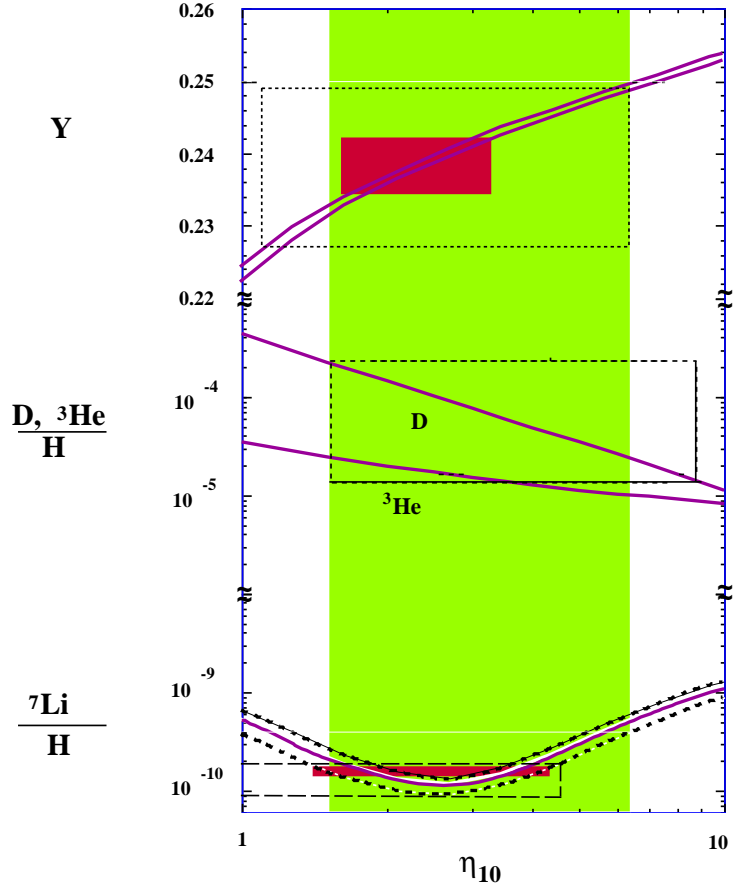}
\end{center}
\caption[]{The light element abundances from big bang
nucleosynthesis as a function of $\eta_{10}$.}
\label{nuc8}
\end{figure}

The dominant product of big bang nucleosynthesis is \he4, resulting in an
abundance of close to 25\% by mass. Lesser amounts of the other light
elements are produced: D and \he3 at the level of about $10^{-5}$ by
number,  and \li7 at the level of $10^{-10}$ by number. The resulting
abundances of the light elements are shown in Figure \ref{nuc8}, over the
range in
$\eta_{10} = 10^{10} \eta$ between 1 and 10.  The curves for the \he4 mass
fraction,
$Y$, bracket the computed range based mainly on the uncertainty of the neutron
mean-life. 
 Uncertainties in the produced \li7 
abundances have been adopted from the results in Hata et al.\cite{hata1}. 
Uncertainties in D and
\he3 production are small on the scale of this figure. 
The dark shaded boxes correspond
to the observed abundances and will be discussed below.

At present, there is a general concordance between the theoretical predictions and
the observational data, particularly, for \he4 and \li7\cite{fo}.  These two
elements indicate that $\eta$ lies in the range $1.55 < \eta < 4.45$.  There is
limited agreement for D/H as well, as will be discussed below. High D/H narrows the
range to $1.5 < \eta < 3.4$ and low D/H is compatible at the $2 \sigma$ level in the
range $4.2 < \eta < 5.3$.

\section{Data}

\subsection {\he4}

\begin{figure}[t]
\hspace{0.5truecm}
\centering
\includegraphics[width=25pc]{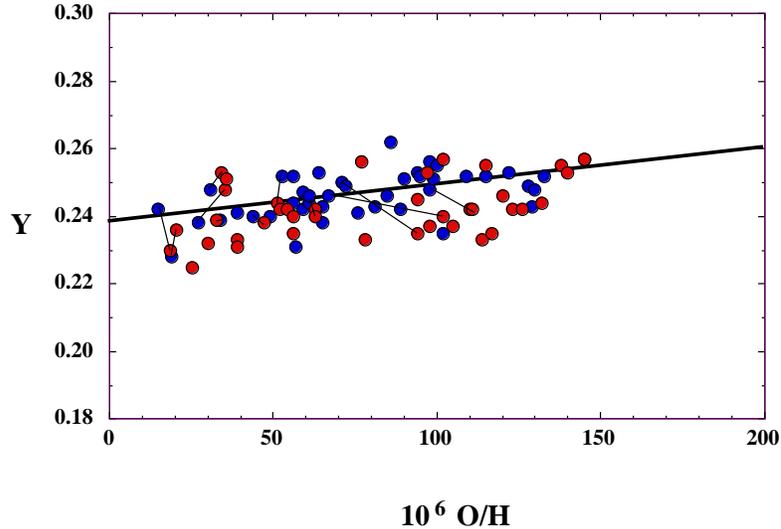}
\caption{{The Helium (Y) and Oxygen (O/H) abundances 
 in extragalactic HII regions, 
from refs. \protect\cite{p}  and from ref.
\protect\cite{it}.   Lines connect the  same regions
observed by different groups. The regression shown leads to the primordial
\he4 abundance given in Eq. (\protect\ref{he4}). }}
\label{he2}
\end{figure}

The primordial \he4 abundance is best
determined from observations  of HeII
$\rightarrow$ HeI recombination lines in extragalactic HII 
(ionized hydrogen) regions.
There is a good collection of abundance information on the \he4 mass
fraction, $Y$, O/H, and N/H in over 70 such regions\cite{p,iz,it}. 
Since \he4 is produced in stars along with heavier elements such as Oxygen,
it is then expected that the primordial abundance of \he4 can be determined
from the intercept of the correlation between $Y$ and O/H, namely $Y_p =
Y({\rm O/H} \to 0)$.   A detailed analysis \cite{fdo2} of the data found
\beq
Y_p = 0.238 \pm 0.002 \pm 0.005
\label{he4}
\eeq
The first uncertainty is purely statistical and the second uncertainty is
an estimate of the systematic uncertainty in the primordial abundance
determination.  The solid box for
\he4 in Figure \ref{nuc8} represents the range (at 2$\sigma_{\rm stat}$)
from (\ref{he4}). The dashed box extends this by including the systematic
uncertainty. The He data is shown in Figure \ref{he2}.

The helium abundance used to derive (\ref{he4}) was determined
using assumed electron densities $n$ in the HII regions obtained from SII data. 
Izotov, Thuan,  \& Lipovetsky \cite{iz} proposed a method based on several He
emission lines to ``self-consistently" determine the electron density.  Their data
using this method yields a higher primordial value
\beq
Y_p = 0.244 \pm 0.002 \pm 0.005
\label{he42}
\eeq

As one can see, the resulting primordial \he4 abundance shows significant sensitivity
to the {\em method} of abundance determination, leading one to conclude that the
systematic uncertainty (which is already dominant) may be underestimated.
Indeed, the determination (\ref{he4})  of the primordial
abundance above is based on a combination of the data in refs. \cite{p}, which alone
yield $Y_p = 0.228
\pm 0.005$, and the data of ref. \cite{it} (based on SII densities) which give
$0.239
\pm 0.002$.  The abundance (\ref{he42}) is based solely on the self-consistent
method yields and the data of \cite{it}. One should also note that a recent
determination
\cite{piem} of the \he4 abundance in a single object (the SMC) also using the self
consistent method gives a primordial abundance of 0.234 $\pm$ 0.003 (actually, they
observe $Y = 0.240 \pm 0.002$ at [O/H] = -0.8, where [O/H] refers to the log of the
Oxygen abundance relative to the solar value, in the units used in Figure \ref{he2},
this corresponds to $10^6$O/H = 135). Therefore, it will useful to discuss some of the
key sources of the uncertainties in the He abundance determinations and prospects for
improvement. To this end, I will briefly discuss, the importance of reddening and
underlying absorption in the H line line measurements, Monte Carlo methods for both H
and He, and underlying absorption in He.

The He abundance is typically quoted relative to H, e.g., He line strengths are
measured relative to 
$H\beta$.  The H data must first be corrected
for underlying absorption and reddening. 
Beginning with an observed line flux $F(\lambda)$, and an equivalent width
$W(\lambda)$, we can parameterize the correction for underlying stellar absorption as
\beq
X_A(\lambda) = F(\lambda) ({W(\lambda) + a) \over W(\lambda)})
\eeq
The parameter $a$ is expected to be relatively insensitive to 
wavelength.
A reddening correction is applied to determine the
intrinsic line intensity $I(\lambda)$ relative to $H\beta$
\beq
X_R(\lambda) = {I(\lambda) \over I(H\beta)} = {X_A(\lambda) \over X_A (H\beta)}
10^{f(\lambda) C(H\beta)}
\label{chb}
\eeq
where $f(\lambda)$ represents an assumed universal reddening law and $C(H\beta$) is
the correction factor to be determined. By minimizing the differences between
$X_R(\lambda)$ to theoretical values,
$X_T(\lambda)$, for $\lambda = H\alpha, H\gamma$ and $H\delta$, one can determine the
parameters
$a$ and $C(H\beta)$ self consistently \cite{osk}, and run a Monte
Carlo over the input data to test the robustness of the solution and to
determine the systematic uncertainty associated with these corrections. 

In Figure \ref{chb} \cite{osk},  the result of such a Monte-Carlo
based on synthetic data with an assumed correction of 2 \AA\ for underlying
absorption and a value for
$C(H\beta) = 0.1$ is shown.  The synthetic data were assumed to have an intrinsic 
2\% uncertainty. While the mean value of the Monte-Carlo results very
accurately reproduces the input parameters, the
spread in the values for $a$ and $C(H\beta)$ are considerably larger than
one would have derived from the direct $\chi^2$ minimization solution due to the
covariance in
$a$ and
$C(H\beta)$. 

\begin{figure}[h] 
\begin{center}
\includegraphics[width=25pc]{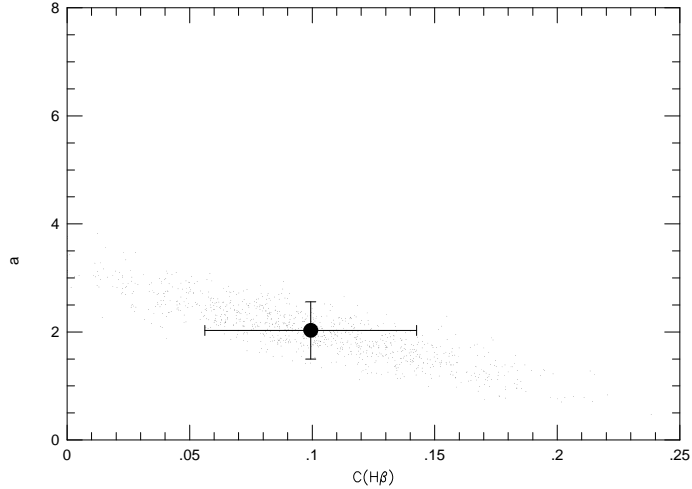}
\end{center}
\caption{A Monte Carlo determination of the underlying absorption $a$ (in \AA\ ), and
reddening  parameter $C(H\beta)$, based on synthetic data.}
\label{chb}
\end{figure}

The uncertainties found for $H\beta$ must next be propagated into  the
analysis for
\he4.
We can quantify the contribution to the overall He abundance
uncertainty due to the reddening correction by propagating the error in
eq. (\ref{chb}). Ignoring all other uncertainties in
$X_R(\lambda) = I(\lambda)/I(H\beta)$, we would write 
\beq
{\sigma_X \over X} = \ln 10~f(\lambda)~\sigma_{C(H\beta)}
\label{echb}
\eeq
In the example discussed above, $\sigma_{C(H\beta)} \sim 0.04$ (from the
Monte Carlo), and values of $f$ are 0.237, 0.208, 0.109, -0.225, -0.345,
-0.396, for He lines at $\lambda \lambda$3889, 4026, 4471, 5876,
6678, 7065, respectively. For the bluer lines, this correction alone is 1
-- 2 \% and must be added in quadrature to any other observational errors
in $X_R$. For the redder lines, this uncertainty is 3 -- 4 \%.  This
represents the {\em minimum} uncertainty which must be included in the
individual He I emission line strengths relative to H$\beta$.

Next one can perform an analogous procedure to that described above
to determine the \he4 abundance \cite{osk}. We
again start with a set of observed quantities: line intensities
$I(\lambda)$ which include the reddening correction previously determined along with
its associated uncertainty which includes the uncertainties in $C(H\beta)$;
the equivalent width
$W(\lambda)$; and temperature $t$. The Helium line intensities are scaled to $H\beta$
and the singly ionized helium abundance is given by
\beq
y^+(\lambda) = {I(\lambda) \over I(H\beta)} {E(H\beta) \over
E(\lambda)} ({W(\lambda) + a_{HeI}
\over W(\lambda)}) {1\over (1+\gamma)} {1\over f(\tau)}
\label{y+}
\eeq
where $E(\lambda)/E(H\beta)$ is the theoretical emissivity scaled to $H\beta$.
The expression (\ref{y+}) also contains a correction factor for underlying
stellar absorption, parameterized now by $a_{HeI}$, a density dependent collisional
correction factor, $(1+\gamma)^{-1}$, and a flourecence correction which depends on
the optical depth $\tau$.  Thus $y^+$ implicitly depends on 3 unknowns, the
electron density, $n$, $a_{HeI}$, and $\tau$. 

One can use 3-6 lines to determine the
weighted average helium abundance, $\bar y$.
From ${\bar y}$, we can calculate the $\chi^2$ deviation from the average,
and minimize $\chi^2$, to determine 
$n, a_{HeI}$, and $\tau$.  Uncertainties in the output parameters are also
determined. 
In principle, under the assumption of small values for the optical
depth $\tau$(3889), it is possible to use only the three bright lines
$\lambda$4471, $\lambda$5876, and $\lambda$6678 and still solve 
self-consistently for He/H, density, and $a_{HeI}$.
Of course, because these lines have relatively low sensitivities 
to collisional enhancement, the derived uncertainties 
in density will be large.

The
addition of $\lambda$7065 was proposed \cite{iz} as a density diagnostic
and then, $\lambda$3889 was later added to estimate the radiative
transfer effects (since these are important for $\lambda$7065).
Thus the five line method has the potential of self-consistently
determining the density and optical depth in the addition to 
the \he4 abundance.
The procedure described here differs somewhat from that proposed in \cite{iz}, in that
the
$\chi^2$ above is based on a straight weighted average, where as in \cite{iz}
the difference of a ratio of He abundances (to one wavelength, say
$\lambda$4471) to the theoretical ratio is minimized. When the reference line is
particularly sensitive to a systematic effect such as underlying stellar absorption,
this uncertainty propagates to all lines this way. 

Adding $\lambda$4026 as a diagnostic line increases the leverage
on detecting underlying stellar absorption.  This is because
the $\lambda$4026 line is a relatively weak line. However, this also
requires that the input spectrum is a very high quality one.
$\lambda$4026 is also provides exceptional leverage to underlying
stellar absorption because it is a singlet line and therefore
has very low sensitivity to collisional enhancement (i.e., $n$)
and optical depth (i.e., $\tau$(3889)) effects.

\begin{figure}[ht] 
\begin{center}
\includegraphics[width=28pc]{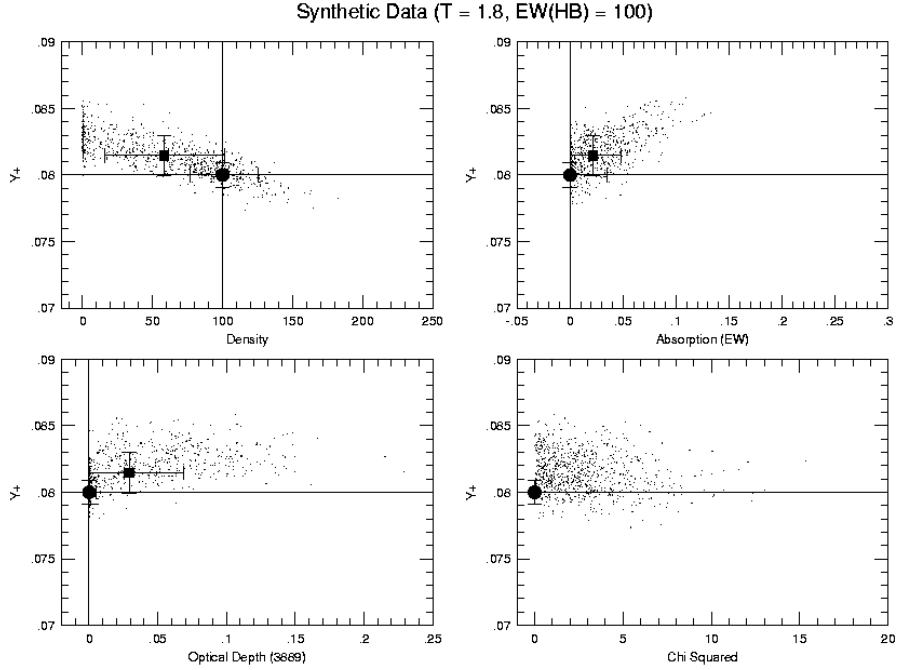}
\end{center}
\caption{Results of modeling of 6 synthetic He~I
line observations.  The 
four panels show the results of a density = 100 cm$^{-3}$,
$a_{HeI}$ = 0, and $\tau$(3889) $=$ 0 model. 
}
\label{mcy1}
\end{figure}

As in the case of the hydrogen lines, Monte-Carlo
simulation of the He data can be used to test the robustness of the solution for
$n,a_{HeI}$,  and $\tau$ \cite{osk}.
Figure \ref{mcy1} presents the 
results of modeling of 6 synthetic He I
line observations.  The  
four panels show the results of a density = 100 cm$^{-3}$,
$a_{HeI}$ = 0, and $\tau$(3889) $=$ 0 model. 
The solid lines show the input values (e.g., He/H = 0.080)
for the original calculated spectrum.  The solid circles
(with error bars) show
the results of the $\chi^2$ minimization solution 
(with calculated  errors) for
the original synthetic input spectrum.  
The small points show the results of Monte Carlo realizations
of the original input spectrum.
The solid squares (with error bars) show the means and dispersions
of the output values for the $\chi ^2$ minimization solutions of
the Monte Carlo realizations.

Figure \ref{mcy1} demonstrates several important points.  First,
the $\chi^2$ minimization solution finds the correct input
parameters with errors in He/H of about 1\% (less than the
2\% errors assumed on the input data, showing the power of
using multiple lines).  
There is a systematic trend for the
Monte Carlo realizations to tend toward higher values of
He/H.  This is because, the inclusion of errors has allowed
minimizations which find lower values of the density and
non-zero values of underlying absorption and optical depth.
 Note that the size of
the error bars in He/H have expanded by roughly 50\% as a result.  We can
conclude from this that simply adding additional lines or physical
parameters in the  minimization does not necessarily lead to the correct
results. In order to use the minimization routines effectively, one must
understand the role of the interdependencies of the individual
lines on the different physical parameters.  Here we have shown
that trade-offs in underlying absorption and optical depth allow
for good solutions at densities which are too low and resulting
in helium abundance determinations which are too high.
Note that in the lower right panel of
Figure \ref{mcy1} that the values of the $\chi ^2$ do not correlate 
with the values of $y^+$.  The solutions at higher values of 
absorption and $y^+$ are equally valid as those at lower
absorption and $y^+$.

\begin{figure}[h] 
\begin{center}
\includegraphics[width=28pc]{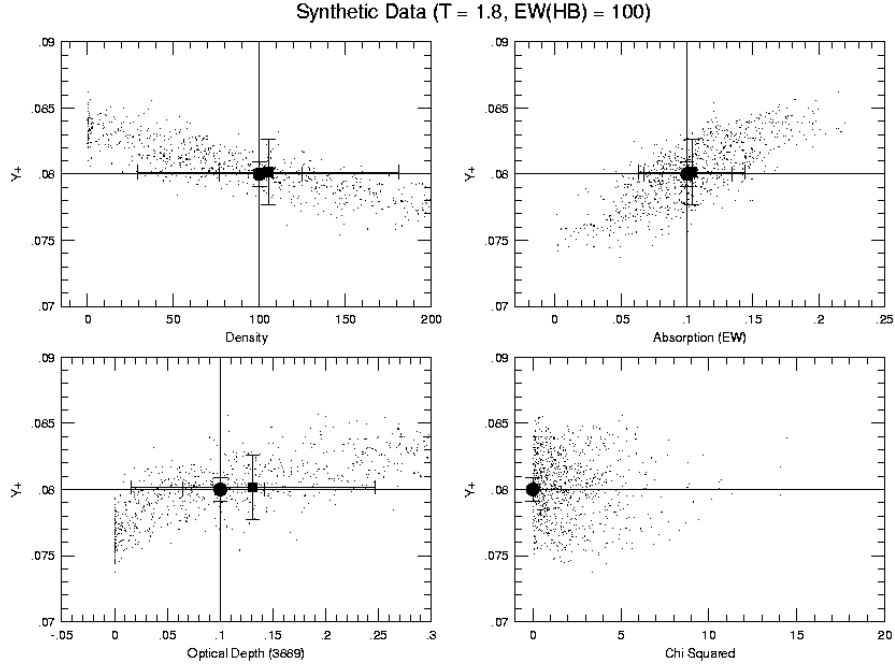}
\end{center}
\caption{Similar plot to Figure \protect\ref{mcy1} except that the
underlying absorption is 0.1 \AA\ and $\tau$(3889) $=$ 0.1. }
\label{mcy2}
\end{figure}

Figure \ref{mcy2}
shows the results of the Monte Carlo when both $\tau$ and $a_{HeI} \ne
0$, and $n = 100$ cm$^{-3}$.
It is encouraging that in perhaps more realistic cases where the input
parameters are non-zero, we are able to derive results very close to their
correct values.  The average of Monte Carlo realizations is remarkably
close to the straight minimization for all of the derived parameters ($n,
a_{HeI}, \tau$ and $y^+$). However, there is an enormous dispersion in
these results due to the degeneracy in the solutions with respect to the
physical input parameters.  This results in error estimates for
parameters which are significantly larger than in the straight
minimization.  For example, the uncertainties in both the density and
optical depth are almost a factor of 3 times larger in the Monte Carlo. 
When propagated into the uncertainty in the derived value for the He
abundance, we find that the uncertainty in the Monte Carlo result (which
we argue is a better, not merely more conservative, value) is a factor of
2.5 times the uncertainty obtained from a straight minimization using 6
line He lines.  This amounts to an approximately 4\% uncertainty in the He
abundance, despite the fact that we assumed (in the synthetic data) 2\%
uncertainties in the input line strengths.   This is an unavoidable
consequence of the method - the Monte Carlo routine explores the 
degeneracies of the solutions and reveals the larger errors that
should be associated with the solutions.

\begin{figure}[h] 
\begin{center}
\includegraphics[width=25pc]{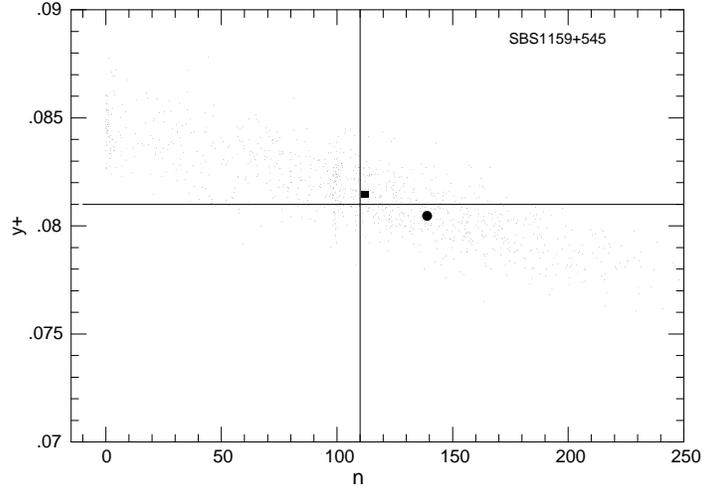}
\end{center}
\caption{A Monte Carlo determination of the helium abundance and electron density
(in cm$^{-3}$) for the region SBS11159+545. Solutions for $a^\prime$ and $\tau$ are not
shown here. }
\label{mcy3}
\end{figure}

 In Figure \ref{mcy3}, I show the result of a single
case based on the data of  ref. \cite{it} for SBS1159+545.
Here, the helium abundance and density solutions are displayed.
The vertical and horizontal lines show the position of the solution in \cite{it}.
The circle shows the position of the our solution to the minimization, and
the square shows the position of the mean of the Monte-Carlo distribution. 
The spread shown here is significantly greater than the uncertainty quoted in
\cite{it}.

\subsection{\li7}

The abundance of \li7 has been determined by observations of over 100
hot, population-II halo stars, and is found to have a very
nearly  uniform abundance\cite{sp}. For
stars with a surface temperature $T > 5500$~K
and a metallicity less than about
1/20th solar (so that effects such as stellar convection may not be
important), the  abundances show little or no dispersion beyond that which
is consistent with the errors of individual measurements.  The Li data from
Ref.\cite{mol}
indicate a mean \li7 abundance of 
\beq
{\rm Li/H = (1.6 \pm 0.1 ) \times 10^{-10}}
\label{li}
\eeq
The small error is statistical and is due to the large number of stars
in which \li7 has been observed. 
 The solid box for \li7 in Figure
\ref{nuc8} represents the 2$\sigma_{\rm stat}$ range from (\ref{li}).

There is, however, an important source of systematic error due to the
possibility that Li has been depleted in these stars, though the lack of
dispersion in the Li data limits the amount of depletion. In fact, a small observed
slope in Li vs Fe and the tiny dispersion about that correlation indicates that
depletion is negligible in these stars \cite{rnb}. Furthermore, the slope may indicate
a lower abundance of Li than that in (6).  The observation\cite{li6o} of the fragile
isotope \li6 is  another good indication that
\li7 has not been destroyed in these stars\cite{li6}.

The weighted mean of the \li7 abundance in the sample of ref. \cite{rnb}
is [Li] = 2.12 ([Li] = $\log$ \li7/H + 12) and is slightly lower than that in eq.
(\ref{li}), the difference is a systematic effect due to analysis methods.  It is
common to test for the presence of a slope in the Li data by fitting a regression of
the form [Li] =
$\alpha +
\beta$ [Fe/H]. These data indicate a rather large slope, $\beta = 0.07 - 0.16$ and
hence a downward shift in the ``primordial" lithium abundance $\Delta$[Li] = $- 0.20
- - 0.09$.  Models of galactic evolution which predict a small slope for [Li] vs.
[Fe/H], can produce a value for $\beta$ in the range 0.04 -- 0.07 \cite{rbofn}. 
Of course, if we would like to extract the primordial \li7 abundance, we must examine the
linear (rather than log) regressions. 
For Li/H = $a^\prime + b^\prime$Fe/Fe$_\odot$, we find $a^\prime = 1 - 1.2 \times
10^{-10}$ and $b^\prime = 40 - 120 \times
10^{-10}$.  A similar result is found fitting Li vs O. 
Overall, when the regression based on the data and other systematic effects are
taken into account a best value for Li/H was found to be \cite{rbofn}
\beq
{\rm Li/H = 1.23  \times 10^{-10}}
\label{li2}
\eeq
with a plausible range between 0.9 -- 1.9 $\times 10^{-10}$. The dashed box in Figure
\ref{nuc8} corresponds to this range in Li/H.

\begin{figure}[h] 
\begin{center}
\includegraphics[width=25pc]{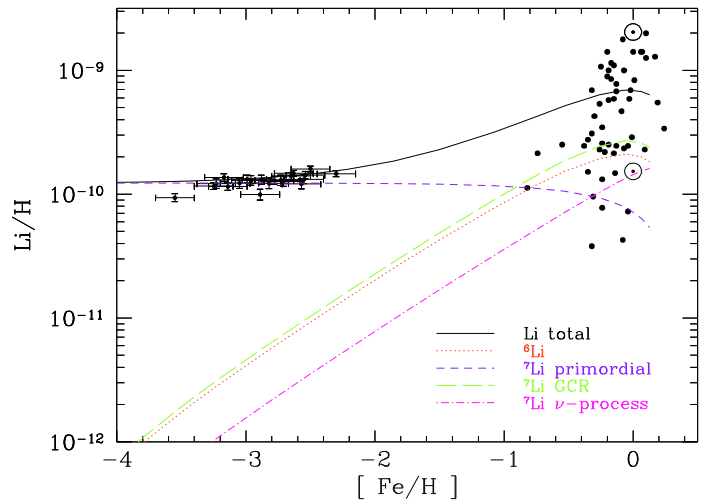}
\end{center}
\caption{Contributions to the total predicted lithium abundance from the adopted GCE
model of ref. \protect\cite{fo98}, compared with low metallicity 
stars (from \cite{rnb}) and a sample of high metallicity stars.
The solid curve is the sum of all  components.  }
\label{li2fig}
\end{figure}

Figure \ref{li2fig} shows the different Li
components for a model with (\li7/H)$_p = 1.23 \times 10^{-10}$. 
The linear slope produced by the model is  $b' = 65\times 10^{-10}$,
and is independent of the input primordial value (unlike the log slope given above).
The model \cite{fo98} includes
in addition to primordial \li7, lithium produced in galactic cosmic ray
nucleosynthesis (primarily $\alpha + \alpha$ fusion), and \li7 produced by
the $\nu$-process during type II supernovae. As one can see, these processes are not
sufficient to  reproduce the population I abundance of \li7, and additional production
sources are needed. 

\section{LiBeB}

The question that one should ask with regard to the above discussion of 
\li7 (and as we will see below when discussing concordance, \li7 will play an 
important role in determining the baryon density $\eta$), is whether or not 
there is additional evidence for the post big bang production of \li7.
There is in fact evidence in the related observation of the intermediate mass 
elements of \li6, Be and B. While these elements are produced in the big bang
\cite{tsof}, their predicted primordial abundance is far below their observed
abundance, which like \li7 is determined by observations of old metal poor halo 
stars.  Whereas in the range $\eta_{10} = 1.5 - 4.5$, standard BBN predicts abundances
of 
\begin{eqnarray}
\li6/{\rm H} & \approx & (2 - 9) \times 10^{-14} \nonumber \\
\be9/{\rm H} & \approx & (0.04 - 2) \times 10^{-17}  \nonumber \\
\b10/{\rm H} & \approx & (0.5 - 3) \times 10^{-19}  \nonumber \\
\b11/{\rm H} & \approx & (0.02 - 1) \times 10^{-16}   
\end{eqnarray}
the observed abundances found in Pop II
halo stars are: \li6/H $\approx$ few $\times ~10^{-12}$, \be9/H $\sim 1 - 10
\times 10^{-13}$, and B/H $\sim 1 - 10 \times 10^{-12}$. It is generally
recognized that these isotopes are not of primordial origin, but rather have
been produced in the Galaxy, through cosmic-ray nucleosynthesis.

Be and B have been observed in the same pop II stars which
show Li and in particular there are a dozen or so stars in which both Be and \li7
have been observed.  Thus Be (and B though there is still a paucity of
data) can be used as a consistency check on primordial Li. 
Based on the Be abundance found in these stars, 
one can conclude that no more than 10-20\% of 
the \li7 is due to cosmic ray nucleosynthesis leaving the remainder
(the abundance in Eq. (\ref{li2})) as primordial.
This is consistent with the conclusion reached in Ref.\cite{rbofn}.

In principle, we can use the abundance information on the other LiBeB isotopes
to determine the abundance of the associated GCRN produced \li7.
As it turns out, the boron data is problematic for this purpose, as there is
very likely an additional significant source for \b11, namely $\nu$-process
nucleosynthesis in supernovae. 
Using the subset of the data for which Li and Be have been observed in the same stars, 
one can extract the primordial abundance of
\li7 in the context of a given model of GCRN.
For example, a specific GCRN model, predicts the ratio of Li/Be as a function
of [Fe/H].  Under the (plausible) assumption that all of the observed Be is
GCRN produced, the Li/Be ratio would yield the GCRN produced \li7 and could
then be subtracted from each star to give a set of primordial \li7 abundances.
This was done in \cite{os} where it was found that the plateau
was indeed lowered by approximately 0.07 dex.
However, it should be noted that this procedure is extremely model dependent.
The predicted Li/Be ratio in GCRN models was studied extensively in \cite{fos}. It was
found that Li/Be can vary between 10 and
$\sim 300$ depending on the details of the cosmic-ray sources and
propagation--e.g., source spectra shapes, escape pathlength
magnitude and energy dependence, and kinematics.

In contrast, the \li7/\li6 ratio is much better determined and far less model
dependent since both are predominantly produced by $\alpha-\alpha$ fusion
rather than by spallation. The obvious problem however, is the paucity of \li6
data.  As more \li6 data becomes available, it should be possible to obtain a
better understanding of the relative contribution to \li7 from BBN and GCRN.

The associated BeB elements are clearly of importance in determining the primordial
\li7 abundance, since
 Li is  produced
together with Be and B in accelerated particle interactions such as cosmic ray
spallation. However, these production processes are not yet fully understood.
Standard cosmic-ray nucleosynthesis is dominated by
interactions originating from accelerated protons and $\alpha$'s on CNO in the ISM,
and predicts that BeB should be  ``secondary'' versus the spallation targets, giving
$\be{}
\propto {\rm O}^2$.  
However, this simple model was challenged by the
observations of BeB abundances in Pop II stars,
and particularly the BeB trends versus metallicity.
Measurements showed that both
Be and B vary roughly {\em linearly} with Fe,
a so-called ``primary'' scaling.
If O and Fe are co-produced (i.e., if O/Fe is constant at low metallicity)
then the data clearly contradicts the canonical theory, i.e. BeB production via
standard GCR's.

There is growing evidence that the
O/Fe ratio is {\em not} constant at low metallicity\cite{isr}, but rather increases
towards low metallicity. This trend offers a solution 
to resolve discrepancy between the observed BeB
abundances as a function of metallicity and the predicted secondary trend
of GCR spallation~\cite{fo98}. As noted above, standard GCR nucleosynthesis
predicts $\be{}\propto {\rm O}^2$, while
observations show $\be{} \sim {\rm Fe}$, roughly; 
these two trends can be consistent if O/Fe is not constant
in Pop II.   A combination of standard GCR nucleosynthesis, and
$\nu$-process production of \b11 may be consistent with current data. 

\begin{figure}[h] 
\begin{center}
\includegraphics[width=25pc]{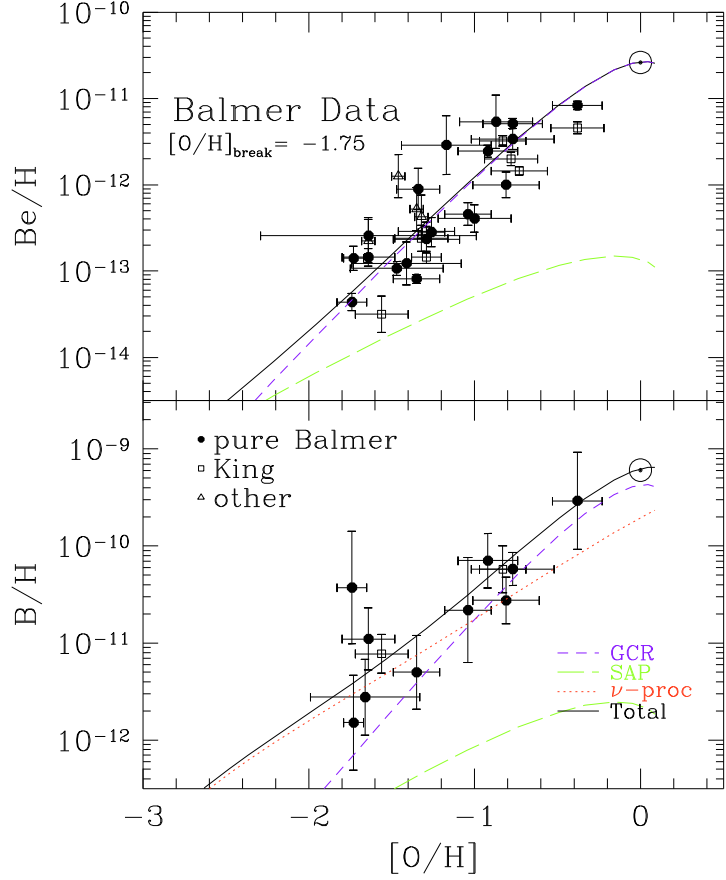}
\end{center}
\caption{Be {\it vs} O ({\it top panel}) and B {\it vs} O ({\it bottom panel}).  
Data shown are found to have a break
point as indicated.  Models are adjusted to achieve the
break point and O/Fe slope of these data.}
\label{fovcfig}
\end{figure}

Thus the nature of the production mechanism for BeB (primary vs. secondary)
rests with the determination of ratio of O/Fe at low metallicity.
In any case, it is clear that given a primary mechanism, it will be dominant
in the early phases of the Galaxy, and secondary mechanisms will dominate
in the latter stage of galactic evolution.  The cross over or break point is
uncertain. In Figure \ref{fovcfig}, a plausible model for the evolution of BeB is shown
and compared with the data \cite{fovc}.  Shown by the short dashed lines are standard
galactic  cosmic-ray nucleosynthesis, which is mostly secondary, but contains some
primary production as well.  The long dashed curves are purely primary, and in the
case of  boron, the $\nu$ process has been included and this too is primary.
The solid curves represent the total Be and B abundance as a function of [O/H].
As one can see such a model fits the data quite well.  

\section{Concordance}

\begin{figure}[h] 
\begin{center}
\includegraphics[width=25pc]{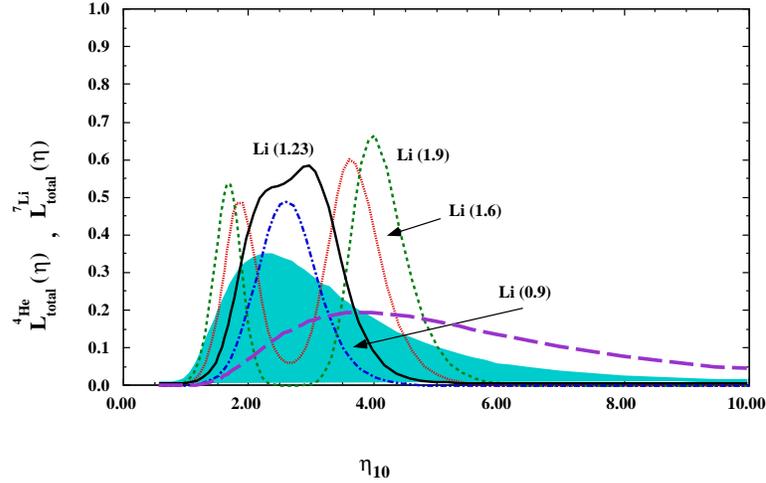}
\end{center}
\caption{Likelihood distributions for four values of primordial $^7$Li/H
($10^{10}\times$ \li7 = 1.9 ({\it dashed}), 1.6 ({\it dotted}), 1.23 ({\it
solid}), and 0.9 ({\it dash-dotted})), and for $^4$He ({\it shaded}) for which we
adopt $Y_p = 0.238\pm 0.002\pm 0.005$ (Eq. (1)). Also shown by the long dashed curve
is the likelihood function based on the \he4 abundance from Eq. (2).}
\label{like1}
\end{figure}

Let us now to turn to the question of concordance between the BBN predictions and the 
observations discussed above.
This is best
summarized in a comparison of likelihood functions as a function of the one free
parameter of BBN, namely the baryon-to-photon ratio $\eta$.  By combining the
theoretical predictions (and their uncertainties) with the observationally determined
abundances discussed above, we can produce individual likelihood functions \cite{fo}
which are shown in Figure \ref{like1}. A range of primordial \li7 values are chosen
based on the the abundances in Eqs. (\ref{li}) and (\ref{li2}) as well as a higher and
lower value. The double peaked nature of the
\li7 likelihood functions is due to the presence of a minimum in the
 predicted lithium abundance in the expected range for $\eta$.  For a given observed
value of \li7, there are two likely values of $\eta$. As the lithium abundance is
lowered, one tends toward the minimum of the BBN prediction, and the two peaks merge. 
Also shown are both values of the primordial \he4 abundances discussed above. 
As one can see, at this level there is clearly concordance between \he4, \li7 and BBN.

\begin{figure}[h] 
\begin{center}
\includegraphics[width=25pc]{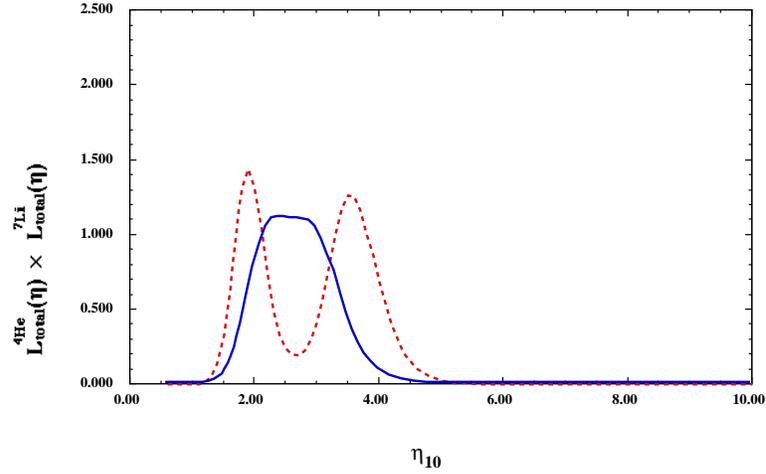}
\end{center}
\caption{Combined likelihood distributions for two values of primordial $^7$Li/H
($10^{10}\times$ \li7 = 1.6 ({\it dashed}), 1.23 ({\it
solid})), and  $^4$He with $Y_p = 0.238\pm 0.002\pm 0.005$ (Eq. (1)). }
\label{like2}
\end{figure}

The combined likelihood, for fitting both elements simultaneously,
is given by the product of two of the functions in Figure \ref{like1}. 
The combined likelihood is shown in Figure \ref{like2}, for the two primordial values
of \li7 in Eqs. (\ref{li}) and (\ref{li2}).  For
\li7$_p = 1.6 \times 10^{-10}$ (shown as the dashed curve), the 95\% CL region covers
the range
$1.55 < \eta_{10} < 4.45$, with the two peaks occurring at
$\eta_{10} = 1.9$ and 3.5. This range corresponds to values of
$\Omega_B$ between
\beq
0.006 < \Omega_B h^2 < .016
\label{omega}
\eeq 
For \li7$_p = 1.23 \times 10^{-10}$ (shown as the solid curve), the 95\% CL region
covers the range
$1.75 < \eta_{10} < 3.90$. In this case, the primordial value is low enough that the
two lithium peaks are more or less merged as is the total likelihood function
giving one broad peak centered at $\eta_{10} \simeq 2.5$. The corresponding 
values of $\Omega_B$ in this case are between
\beq
0.006 < \Omega_B h^2 < .014
\label{omega2}
\eeq

When deuterium is folded into the mix, the situation becomes more complicated.
Although there are several good measurements of deuterium in quasar absorption systems
\cite{burles}, and many of them giving a low value of D/H $\simeq (3.4 \pm 0.3) \times
10^{-5}$ \cite{bty}, there remains an observation with D/H nearly an order of magnitude
higher D/H $\simeq (2.0 \pm 0.5) \times 10^{-4}$ \cite{webb}. 

\begin{figure}[h] 
\begin{center}
\includegraphics[width=25pc]{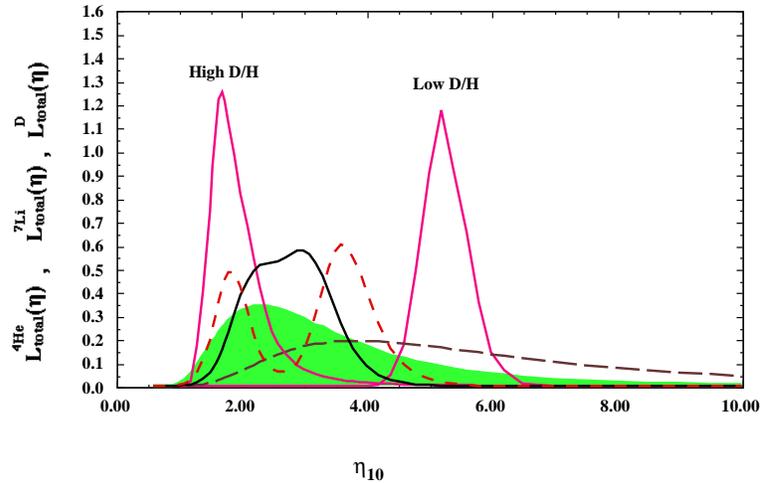}
\end{center}
\caption{Likelihood distributions for two values of primordial $^7$Li/H
($10^{10}\times$ \li7 = 1.6 ({\it dashed}) and 1.23 ({\it
solid})), and  $^4$He with $Y_p = 0.238\pm 0.002\pm 0.005$  from Eq.
(\protect\ref{he4}) (shaded) and $Y_p = 0.244\pm 0.002\pm 0.005$  from Eq.
(\protect\ref{he42}) (long dashed). Also shown are the two likelihood functions for
high and low D/H as marked.}
\label{like3}
\end{figure}

Because there are no known astrophysical sites for the production of
deuterium, all observed D is assumed to be primordial. As a result,
any firm determination of a deuterium abundance establishes an upper bound
on $\eta$ which is robust.  Thus the ISM measurements\cite{lin} of D/H = 1.6 $\times
10^{-5}$ imply an upper bound $\eta_{10} < 9$.

 It is interesting to
compare the results from the likelihood functions of \he4 and \li7 with
that of D/H.  This comparison is shown in Figure \ref{like3}.
 Using the higher value of D/H = $(2.0 \pm 0.5) \times 10^{-4}$, we would find 
excellent agreement between \he4, \li7 and D/H.  The predicted range for $\eta$ now
becomes
\beq
1.6 < \eta_{10}  < 3.2
\label{finh}
\eeq
with the peak likelihood value at $\eta_{10} = 2.1$, \he4 and \li7 abundances from 
eqs. (\ref{he4}) and (\ref{li2}) respectively. 
This  corresponds to $\Omega_B
h^2 = .008^{+.004}_{-.002}$.  
The higher \li7 abundance of eq. (\ref{li}) drops the peak value down slightly
to $\eta_{10} = 1.8$ and broadens the range to 1.5 -- 3.4.
The higher \he4 abundance shifts the peak and range (relative to eq. (\ref{finh}))
up to 2.2 and 1.7 -- 3.5.

\begin{figure}[h]
\hspace{0.5truecm}
\centering
\includegraphics[width=25pc]{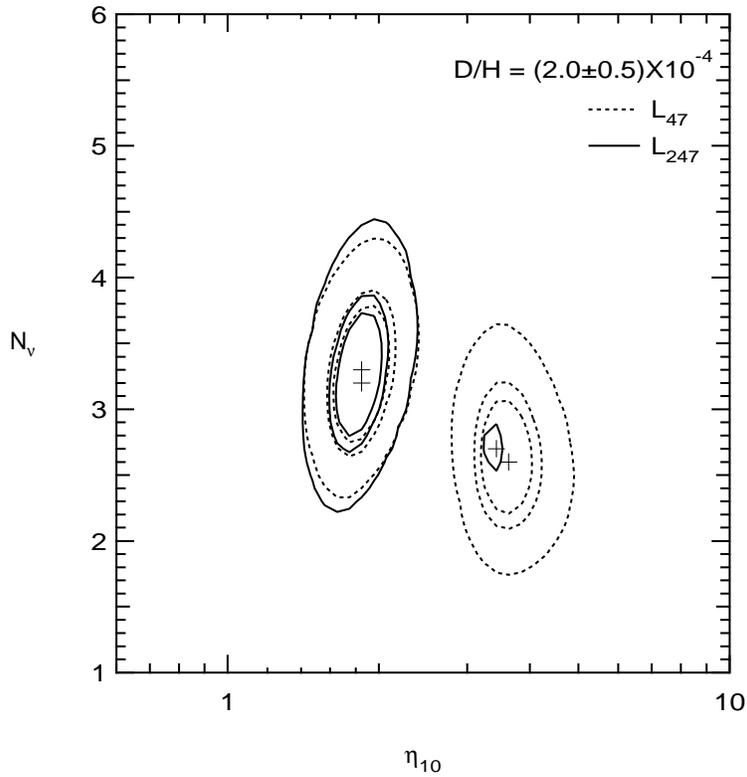}
\caption{50\%, 68\% \& 95\% C.L. contours of $L_{47}$ and
                 $L_{247}$ where observed abundances are given by
                 eqs. (\protect\ref{he4} and  \protect\ref{li}), and
high D/H.}
\label{fig:oth34}
\end{figure}
 
 If instead, we assume that the low value 
of D/H = $(3.4 \pm 0.3) \times 10^{-5}$ \cite{bty} is the primordial abundance,
 there  is hardly any overlap between the D and \li7, particularly for the 
lower value of \li7 from eq. (\ref{li2}).  There is also very limited overlap
between D/H
and \he4, though because of the flatness of the \he4 abundance with respect to
$\eta$, as one can see, the likelihood function for the larger value of \he4
from eq. (\ref{he2}) is very broad.   In this case, D/H is just compatible (at the 2
$\sigma$ level) with the other light elements, and the peak of the likelihood function
occurs at  roughly $\eta_{10} = 4.8$ and with a range of 4.2 -- 5.6.

\begin{figure}[htbp]
\hspace{0.5truecm}
\centering
\includegraphics[width=25pc]{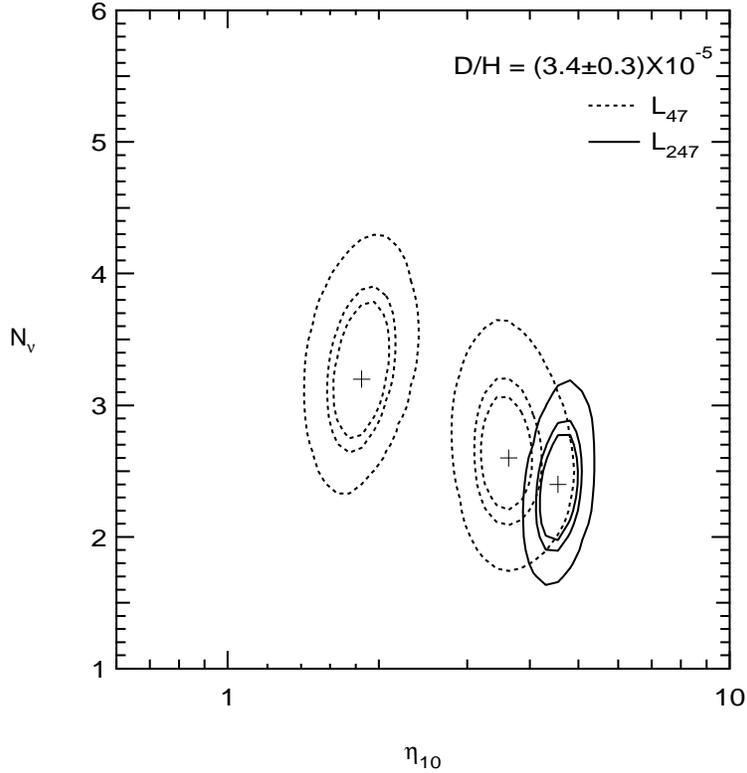}
\caption{50\%, 68\% \& 95\% C.L. contours of $L_{47}$ and
                 $L_{247}$ where observed abundances are given by
                  eqs. (\protect\ref{he4} and  \protect\ref{li}), and
low D/H.}
\label{fig:oth35}
\end{figure}

It is important to recall however, that the true uncertainty in the low
D/H systems might be somewhat larger.  Mesoturbulence effects\cite{lev} allow D/H to
be as large as $5 \times 10^{-5}$. In this case, the peak of the D/H likelihood
function shifts down to 
$\eta_{10} \simeq 4$, and there would be a near perfect overlap
with the high $\eta$ \li7 peak and since the \he4 distribution function is
very broad, this would be a highly compatible solution.

We can obtain still more information regarding the compatibility of the observed
abundance and BBN by considering generalized likelihood functions where we allow
$N_\nu$ to vary as well \cite{fo,oth2,oth3,sark2}. The likelihood functions now 
become functions of two parameters ${\cal L}(\eta,N_\nu)$.

 The peaks of the distribution as
well as the allowed ranges of $\eta$ and $N_\nu$ are  
easily discerned in the 
contour plots of Figures \ref{fig:oth34} and \ref{fig:oth35} which show
the 50\%, 68\% and 95\% confidence level contours in $L_{47}$ and
$L_{247}$ projected onto the $\eta$--$N_\nu$ plane, for high and low D/H as
indicated.  $L_{47}$ corresponds to the likelihood function based on \he4 and \li7
only, whereas $L_{247}$ includes D/H as well.  The crosses show the location of the 
peaks of the likelihood functions.
$L_{47}$ peaks at $N_\nu=3.2$, $\eta_{10}=1.85$  and at $N_\nu=2.6$,
$\eta_{10}=3.6$.  The 95\% confidence level allows the following ranges
in $\eta$ and $N_\nu$
\beq
1.7\le N_\nu\le4.3  \qquad \qquad
1.4\le\eta_{10}\le 4.9 
\eeq
Note however that the ranges in $\eta$ and $N_\nu$ are strongly
correlated as is evident in Figure \ref{fig:oth34}.

With high D/H, $L_{247}$ 
peaks at $N_\nu=3.3$, and also at $\eta_{10}=1.85$. 
In this case
the 95\% contour gives the ranges
\beq
2.2\le N_\nu\le4.4 \qquad \qquad
1.4\le\eta_{10}\le 2.4
\eeq
Note that within the 95\% CL range, there is also a small area 
with $\eta_{10} = 3.2 - 3.5$ and $N_\nu = 2.5-2.9$.

Similarly, for  low D/H, $L_{247}$ 
peaks at $N_\nu=2.4$, and $\eta_{10}=4.55$. 
The 95\% CL upper limit is now $N_\nu < 3.2$, and the range for 
$\eta$ is $ 3.9 < \eta_{10} < 5.4$.  It is important to stress that these abundances
are now consistent with the standard model value of $N_\nu = 3$ at the 2 $\sigma$
level.

\section*{Acknowledgments}
This work was supported in part by
DoE grant DE-FG02-94ER-40823 at the University of Minnesota.

%

\end{document}